\documentclass[preprint]{epl}
\usepackage{isolatin1,epsfig,psfrag}

\newcommand{\x}{{\bf x}}
\newcommand{\average}[1]{\overline{#1}}
\newcommand{\xb}{\bar{x}}
\newcommand{\xs}{x^*}

\title{Evolutionary games and quasispecies}
\shorttitle{Evolutionary games}
\shortauthor{M. Lässig \etal}
\author{Michael Lässig\inst{1}\thanks{To 
whom correspondence should be addressed.
E-mail: \texttt{lassig@thp.uni-koeln.de}} \and Francesca Tria\inst{2}
\and Luca Peliti\inst{2}\thanks{Associato INFN, Sezione di Napoli.}}
\institute{
  \inst{1} Institut für theoretische Physik, Universität zu Köln,
Zülpicher Str.\ 77, D--50937 Köln (Germany)\\
  \inst{2} Dipartimento di Scienze Fisiche and Unità INFM,
Università ``Federico II'', Complesso Monte S. Angelo,
I--80126 Napoli (Italy)
}
\begin{document}
\maketitle
\begin{abstract}
We discuss a population of sequences subject to mutations
and frequency-dependent selection, where the fitness of a
sequence depends on the composition of the entire population. 
This type of dynamics is crucial to understand, for example,
 the {\em coupled} evolution of different strands in a viral
 population. 
Mathematically, it takes the form of
a reaction-diffusion problem that is nonlinear
in the population state. In our model system, the fitness
is determined by a simple mathematical game, the hawk-dove
game. The stationary population distribution is found to
be a quasispecies with properties different from those which
hold in fixed fitness landscapes.
\end{abstract}
\section{Introduction}
The roles of chance and determinism are a central
theme in evolutionary biology. Eigen's \textit{quasispecies
theory}~\cite{Eigen} has been pivotal as a simple quantitative
model for the intertwined effects of
random mutations and Darwinian selection. These
forces act on individuals with a \textit{genotype} defined
by an $L$-letter sequence
$\sigma = (\sigma_1,\dots,\sigma_L)$.
The individuals reproduce at a rate $f(\sigma)$ called the
\textit{fitness} of the genotype $\sigma$,
and are subject to random mutations of
the sequence elements at a rate $\mu$. The population
is  described by a time-dependent frequency distribution
$P(\sigma)$.
The evolution of $P(\sigma)$ is described by a deterministic
equation in the limit of large population size, when sampling
errors in the reproduction-mutation process become
negligible. For small $\mu$, the
evolution is dominated by reproductive success and produces
a {\em quasispecies}, that is, a stationary population
distribution $p_\mathrm{s}(\sigma)$ peaked around
the genotype $\sigma^*$ of maximal fitness. Large mutation
rates, on the other hand, wipe out fitness differences
and lead to a broad distribution. These
two regimes are linked by a crossover which, depending
on the fitness ``landscape'' $f(\sigma)$, may become a phase
transition in the limit of infinite sequence length. The
transition point is called the \textit{error threshold\/}.

Quasispecies evolution can be seen as a reaction-diffusion problem
in sequence space
that is conceptually related to problems in statistical
physics. It takes the form of an (imaginary-time)
Schrödinger equation for the population state,
$\partial_t P = H P$, with $f(\sigma)$ as scalar potential
and the kinetic term describing mutations~\cite{Baake}. This type
of problem is mathematically tractable since $H$ is a
linear operator, and has been studied for many
different fitness landscapes. In particular, there
can be extended subsets of sequence space---called
\textit{neutral networks} $\Gamma_{\x}$---where the
sequences encode the same \textit{phenotype} ${\x}$ and,
therefore, the fitness is constant. The evolution of the
phenotype population
$p({\x}) = \sum_{\sigma \in \Gamma_{\x}} P(\sigma)$
can often be described by a projected equation,
$\partial_t p = H_{\x} p$,
taking into account the varying number of genotypes
coding for the same phenotype.
This means that the selection among different neutral networks
is not only determined by their ``fitness'' (the reproduction rate) but
also by their ``robustness'': i.e., to the probability that a random
mutation leaves a genotype on the same neutral network~\cite{Nimwegen}.
In some cases, this effect can be described by a suitably defined
``mutational entropy''~\cite{Peliti}.

The stationary distribution $p_\mathrm{s}({\x})$ may again
be a quasispecies around the phenotype ${\x}^*$
of maximal fitness, while the sequence population within
each neutral network $\Gamma_{\x}$ remains broad. Examples
are RNA sequences $\sigma$ with only the folding
configuration ${\x}(\sigma)$ determining the fitness.
In this case, the neutral networks $\Gamma_{\x}$ consist of
all sequences with the same fold ${\x}$~\cite{Schuster}.

A step towards a more realistic theory of mutation-selection
processes is to take into account the dependence of the fitness
on the population state. 
This is particularly apparent, for example, in viral 
evolution. The reproductive success of one viral strand
will depend on the other strands that coexist in the
same population. 
At the level of phenotypes,
a well-known way to describe such coupled systems is
\textit{evolutionary game theory}~\cite{MaynardSmith,Weibull,Sigmund,game}.
Here we limit ourselves to the simplest form of a
mathematical game, which is described by
a set of basic \textit{strategies} $i = 1, \dots, s$ and
a matrix $\mathsf{A}=(A_{ij})$
which specifies the relative fitness or \textit{payoff} of
strategy $i$ played against strategy $j$. Then
the phenotype $\x$ is a mixed strategy
$(x_1,\ldots,x_s)$, where $0 \le x_i \le 1$
is the probability that the individual plays
basic strategy $i$.  One can then define the average strategy
$\bar \x = (\xb_1, \dots, \xb_s)$ by averaging the
phenotype over the whole population. 
A (simplified) model of the evolutionary
process is called
\textit{adaptive dynamics}~\cite[ch.~9]{Sigmund}. The
time-dependent population state is approximated by a sequence
of phenotypically homogeneous states, each evolving from
the previous one by invasion of the neighboring mutant
with the highest relative payoff.
It can be shown that
adaptive dynamics leads to strategic optimization:
every stable fixed point $\x^*$ of the population average
is a \textit{Nash equilibrium}~\cite{Nash},
that is, a mixed strategy
that maximizes the payoff against itself (see the more
precise definition below).

In this Letter, we extend the quasispecies approach
to populations evolving according to evolutionary
games. This is appropriate for the example of viral 
evolution, where the frequency-dependent fitness differences
are balanced by mutations (occuring with a high rate 
of $10^{-4}$ per nucleotide per generation).
Unlike for standard quasispecies theory,
the ``Hamiltonian'' $H$ is now a nonlinear operator acting on the
population state. Quite remarkably, the stationary population
distributions $p_\mathrm{s}(\sigma)$ 
and $p_\mathrm{s}(\x)$ can still be calculated
exactly in some cases. We will demonstrate this for sequences
playing the \textit{hawk-dove game}, one of the simplest
games with a nontrivial Nash equilibrium. An individual's
phenotype is associated with a mixed strategy $\x$ uniquely
determined by its genotype $\sigma$. There are extended
neutral networks $\Gamma_{\x}$ consisting of all sequences
encoding the same strategy $\x$. The phenotype population
$p(\x)$ is found to be a quasispecies. The population average
$\bar \x$ deviates from the Nash equilibrium by an amount depending
on the mutation rate. The properties of the quasispecies are
different from those in a fixed fitness landscape. This
reflects the fact that the system exhibits a higher degree
of near-neutrality, as will be discussed in detail below.

\section{Phenotypes and evolutionary game theory}
Consider a population whose phenotypes correspond to the
mixed strategies ${\bf x}$ of a game with payoff matrix
$\mathsf{A}$. That is,
an individual of phenotype ${\bf x}$ plays the basic strategy
$i$ with probability $x_i$ ($i = 1,\dots, s$), and the
payoff for a mixed strategy ${\bf x}$
against a mixed strategy ${\bf x'}$ is assumed to be bilinear:
\begin{equation}
\phi({\x} | {\x'}) = \sum_{i,j=1}^s A_{ij} x_i x'_j \;.
\end{equation}
A Nash equilibrium is defined to be a strategy $\x^*$ that
is optimal against itself, i.e.,
\begin{equation}
\phi(\x|\x^*) \le \phi(\x^*|\x^*) \qquad \mbox{for all
strategies $\x$} \;.
\label{Nash:eq}
\end{equation}
In evolutionary game theory, the game payoff determines
the relative fitness of individuals. The population state is
a time-dependent phenotype distribution $p(\x)$; we suppress
the dependence on $t$ in the notation here and below.
Assuming random mixing of the population, the fitness
of a phenotype $\x$ is given by
\begin{equation}
\int \upd \x' \,\phi(\x | \x')\, p(\x') = \phi(\x|\bar \x)
\;,
\label{phi}
\end{equation}
where $\bar \x$ denotes the average strategy.
Hence, the population  evolves according to
\begin{equation}
\partial_t p(\x) =  [\phi(\x|\bar \x) - \phi(\bar \x|\bar \x) ]\, p(\x)\;;
\label{pdot:eq}
\end{equation}
subtracting the average fitness $\phi(\bar \x|\bar \x)$
ensures that the
normalization of the distribution $p(\x)$ is preserved.
In general, this
equation does not have a unique stationary solution. It is
easy to see that $p(\x)$ is an attractive fixed point
if and only if the population average $\bar \x$ is a Nash
equilibrium and all phenotypes $\x$ in the support of $p$ are
degenerate in fitness, i.e.,
$\phi(\x|\x^*) = \phi(\x^*|\x^*)$.

In this letter, we will focus on the classical hawk-dove
game, which has the two basic strategies ``hawk'' ($i=1$) and
``dove'' ($i=2$); see, e.g.,~\cite[chap.~6]{Sigmund}.
Consider individuals of a population
competing for resources of reproductive value $\lambda$.
Doves avoid confrontation, while hawks escalate fights.
Thus, on average, two doves will share the resources.
A dove meeting a hawk will leave the entire resources
to the hawk. Two hawks will also share the resources
but have to pay a confrontation cost $C \lambda$.
These payoffs can be written in matrix form,
\begin{equation}
\mathsf{A}=\lambda \left(
\begin{array}{cc}(1-C)/2&1\\0&1/2\end{array}\right).
\label{payoffmat:eq}
\end{equation}
An individual with strategy $\x$ plays, by
definition, hawk with probability $x_1\equiv x$ and dove with
probability $x_2 = 1-x$.
The unique Nash equilibrium of the hawk-dove game
is $\xs =\min(1,1/C)$.
In the following, we assume $C > 1$ so that the Nash
equilibrium is a mixed strategy.
According to eqs.~(\ref{phi}) and
(\ref{payoffmat:eq}), the relative fitness of an
arbitrary mixed strategy is given by
\begin{equation}
\phi(\x|\bar \x) - \phi(\bar \x|\bar \x)= 
\frac{\lambda}{2 \xs} (x - \xb)(\xs - \xb) \;.
\label{avpayoff:eq}
\end{equation}

\section{Genotypes and mutations}
The phenotype of an individual is assumed to be uniquely
determined by its genotype. We consider here genotypes
with $L$ loci contributing additively to the phenotype.
Each locus has $c$ alleles, of which $a$
encode the hawk strategy and $c-a$ the dove strategy.
We denote by $q = a/c$ the relative fraction of
hawk coding alleles.
Such genotypes can be represented by sequences
$\sigma = (\sigma_1, \dots, \sigma_L)$ whose letters
take the values 0 and 1 representing dove and hawk
alleles, respectively. The corresponding phenotype
is given by
\begin{equation}
x(\sigma) = \frac{1}{L} \sum_{\alpha = 1}^L \sigma_\alpha\;,
\label{gp}
\end{equation}
and takes the discrete values $x = k/L$, where $k=1,\dots, L$
is the number of hawk alleles.
During a time interval of duration $\upd t$ any given locus
in an individual's genotype mutates with probability $\mu\,\upd t$
into a randomly chosen allele. It is easy to see that in the
absence of selection
the mutations change the discrete population distribution,
\begin{equation}
\partial_t p(x) = - [J(x)-J(x-L^{-1})]\;,
\end{equation}
where $J(x)$ is the net probability current between
all genotypes with $k=Lx$ and with $k+1$ hawk alleles:
\begin{equation}
J(x) = \frac{\mu c}{c-1}
 L [(1-x) q \, p(x) - (x+L^{-1})(1-q) \, p(x+L^{-1})]  \;.
\end{equation}
This neutral evolution leads to a stationary state
$p_0 (x)$ where all alleles are equally probable:
\begin{equation}
p_0(x) = {L\choose Lx} q^{Lx} (1-q)^{L(1-x)} \;.
\label{p0bar}
\end{equation}
Hence the average phenotype in the population is
$\xb_0 = q$.

\section{Quasispecies equation}
In the following, we describe systematically the interplay
between mutations and selection, which leads to nontrivial
stationary population states $p_\mathrm{s}(x)$.
Assuming that the two kinds of processes act in parallel,
we obtain the mutation-selection equation
\begin{equation}
\partial_t p(x) =- \left [ J(x)-J(x-1/L)\right]
+ \frac{\lambda L}{2 \xs} (x - \xb)(\xs - \xb) p(x)
\;.
\label{evoeq:eq}
\end{equation}
This type of dynamics is usually referred to as
\textit{paramuse} models~\cite{Baake}. 
The $L$ dependence
in the fitness has been introduced in order to
obtain a well-behaved large-$L$ limit, i.e.,
a distribution of the form $p(x)\propto 
\exp\left(-L\mathcal{F}(x)\right)$, in
analogy with the thermodynamic limit in statistical mechanics
and with the standard quasispecies theory~\cite{Eigen}.
Since the phenotype average $\bar x$ depends on $p(x)$,
eq.~(\ref{evoeq:eq}) leads to a nonlinear
equation for the stationary population state $p_\mathrm{s}(x)$.

In the present case, the stationary solution can be obtained
exactly. Indeed, one can argue that each different locus undergoes
an independent mutation-selection process, where the
only interaction is encoded in $\xb$. Thus one may
look for a probability distribution that factorizes 
into a product of single-locus probabilities (this
is a general property when the fitness is a linear
functional of the genome~\cite{Rumschitzky}). Since
all loci are equal in the present model, the stationary
distribution $p_\mathrm{s}(x)$ in the presence
of selection is still a binomial, but with a different
average $\xb$:
\begin{equation}
p_\mathrm{s}(x)={L\choose Lx} \xb^{Lx} (1-\xb)^{L(1-x)}.
\label{exact:eq}
\end{equation}
The scaled variance  of this distribution, 
$\sigma^2 \equiv L \, \overline{(x-\xb)^2}$, is linked to  
to $\xb$ by
\begin{equation}
\sigma^2 = \xb (1-\xb)\;.
\label{sigma:eq}
\end{equation}
On the other hand, the equation of motion (\ref{evoeq:eq})
implies a hierarchy of evolution equations for the
moments of $p(x)$. The leading equation of this hierarchy
reads
\begin{equation}
\partial_t \xb = \frac{\mu c}{c-1}(\xb - q)+ \frac{\lambda}{%
2 x^*}\sigma^2\,(\xb - x^*).
\label{shift:eq}
\end{equation}
Inserting eq.~(\ref{sigma:eq}) we obtain a self-consistency
equation for $\xb$, which is of cubic order.
The resulting values of $\xb$ and $\sigma^2$ at stationarity
are plotted in Fig.~1
against the effective mutation rate $\mu/\lambda$
together with the results from numerical simulations
for $L=16$. In accordance with the results obtained above,
there is no $L$-dependence in these quantities (except
for the effects of rounding errors).
\begin{figure}[htb]
\psfrag{mul}[l]{$\mu/\lambda$}
\psfrag{xb}{$\xb$}
\psfrag{sigma2}{$\sigma^2$}
\begin{center}
\includegraphics[width=6cm]{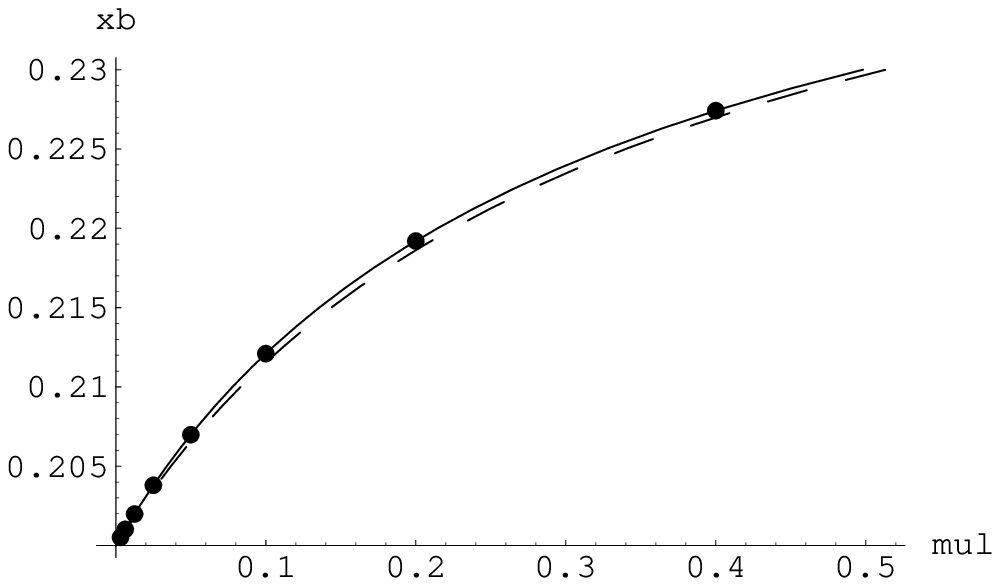}
\hspace{0.2cm}
\includegraphics[width=6cm]{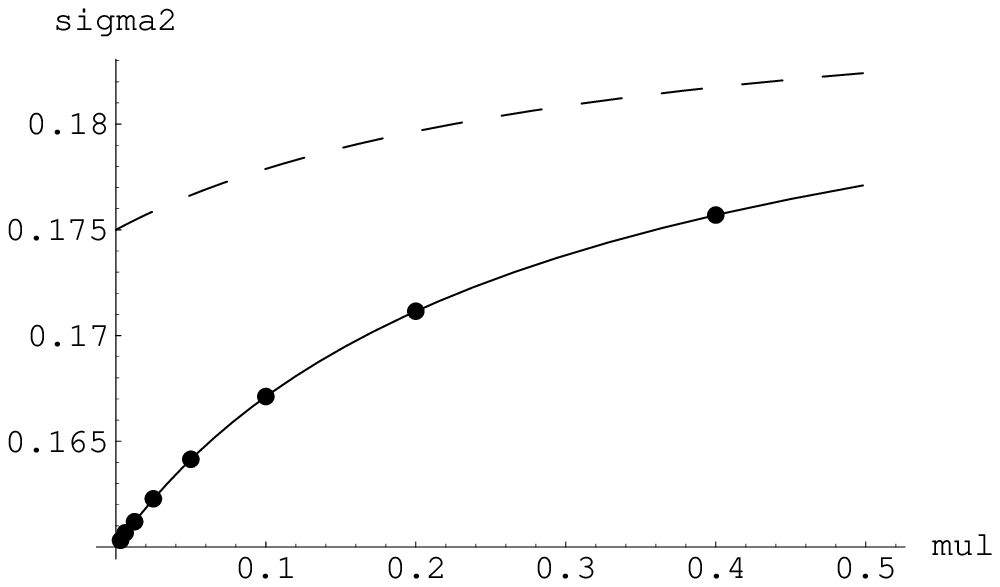}
\end{center}
\caption{
Average strategy $\xb$ (left) and strategy variance
$\sigma^2$ (right) of the phenotypic quasispecies as a function of the
effective mutation rate $\mu/\lambda$. The points are
simulation data for $c=4$, $a=1$, $\lambda=1$, $\xs=0.2$,
$L=16$. The continuous lines give the solutions of 
eqs.~(\ref{sigma:eq},\ref{shift:eq}).
The dashed lines represent the solution obtained
from the $\Omega$-expansion by neglecting the last term
in eq.~(\ref{skewness:eq}).}
\label{fig.1}
\end{figure}

Note that a straightforward application of the well-known
$\Omega$-expansion (see, e.g., \cite[Chap.~IX]{vanKamp})
overestimates the variance of $x$ and therefore
does not give the correct result. This can be seen
from the next-to-leading evolution
equation for the moments of $p(x)$, 
\begin{equation}
\partial_t \sigma^2 =\frac{\mu c}{c-1}
\left[-2 \sigma^2 + \xb (1-2q)+q\right]+
\frac{\lambda}{2 x^*}L^2 (x^*-\xb)\average{
(x-\xb)^3}.
\label{skewness:eq}
\end{equation}
In the $\Omega$-expansion it is assumed that the first
two moments can be obtained self-consistently
from the {\em truncated} system of equations
(\ref{shift:eq}) and (\ref{skewness:eq}), 
neglecting all higher cumulants as $L\to\infty$.
Indeed, the $m$-th order cumulant of $p_\mathrm{s}(x)$ 
scales like $L^{1-m}$ as required by the $\Omega$-expansion.
However, in eq.~(\ref{skewness:eq}) the skewness
is multiplied by a factor $L^2$, so that its contribution
to $\sigma^2$ remains finite in the limit $L \to \infty$
and the truncation fails.

As a function of $\mu/\lambda$, the solution
 $p_\mathrm{s}(x)$  describes
a crossover between two dynamical regimes.
\newline (i) In the {\em fast mutation} regime
$\mu/ \lambda \gg 1$, the
evolution becomes effectively neutral. The phenotype
distribution approaches the asymptotic form $p_0(x)$
given by eq.~(\ref{p0bar}); we have
\begin{equation}
\xb = q + O \left ( \frac{\lambda}{\mu} \right )\;.
\end{equation}
\newline (ii) In the {\em slow mutation} regime
$\mu/ \lambda \ll 1$, the phenotype average approaches
the Nash equilibrium but the variance remains finite,
\begin{equation}
\xb = \xs + \frac{\mu c}{c-1} \frac{2 x^*}{\lambda}(q- \xs)
+o\left(\frac{\mu}{\lambda}\right)\;,
\qquad
\sigma^2 = \xs (1-\xs)+O\left(\frac{\mu}{\lambda}\right)\;.
\end{equation}

The salient feature of this crossover is that the
phenotypic quasispecies is always broad, even in
the limit $\mu \to 0$.
(In contrast, the standard
quasispecies in a  fixed fitness landscape would be localized
around a master phenotype with a variance $\sigma^2 = O(\mu)$.)
This broadness reflects the fact that at a mixed Nash equilibrium,
all participating basic strategies have the same fitness.
Hence, as $\xb$ approaches $\xs$, the fitness differences
between phenotypes become small. 

\section{Discussion} 

A generic feature of this mutation-selection dynamics is broad 
equilibria around Nash points, with a larger degree of sequence
divergence than in many fixed fitness landscapes. 
It is clear that this method can be applied without
major changes to evolutionary games with unconditional strategies, in which
the payoff $\phi(\x|\x')$ is a linear function of $\x'$. The expression of the
mutational entropy can be easily generalized to games with
$s$ pure strategies, encoded by $a_i$ ($i=1,\ldots,s$) alleles respectively.
Systems where different
loci have a different weight in the genotype-phenotype
mapping (\ref{gp}) can be treated along the
lines of ref.~\cite{Peliti}. In general, the stationary distribution
of strategies will be close to an evolutionary stable strategy for small
mutation rate, if there is one.
The situation is different in games with conditional strategies, like
the celebrated Prisoner's Dilemma game~\cite{Axelrod},~\cite[p.~101]{Sigmund}.
In this case, the payoff of a given strategy depends on more properties
of the population than the average strategy alone. 
This case will be the subject of a separate
publication~\cite{LPT}. It is also clear that the present approach
can be generalized to the case of asymmetric games, which can
be used to model the co-evolution of different (but interacting)
populations.

The main limitation of the quasispecies approach is the assumption of
large population sizes. In finite populations, there are
sampling fluctuations,  which lie at the heart of the Neutral
Evolution approach~\cite{Kimura,Ohta}. Recently these
fluctuations have been incorporated into phenotypical
evolutionary game theory, using  a quantum-mechanical
formalism~\cite{Lassig}.
Analogous finite-population effects in our
co-evolutionary sequence dynamics lead to a quantum field
theory, which will also be the subject of future work~\cite{LPT}.

\acknowledgments
We are grateful to Oliver Redner and Joachim Hermisson for pointing
out an error in the preprint version of the manuscript and suggesting
that an exact equation for $\xb$ could be obtained. We also
thank the Centro di Ricerca Matematica ``Ennio De Giorgi''
for hospitality in Pisa during part of this work.

\end{document}